\documentclass[11pt]{article}

\usepackage[a4paper,hmargin=1.72cm,vmargin=2.4cm]{geometry}
\usepackage{amsmath,amssymb,amsthm,mathtools}
\usepackage{bbm}
\usepackage{hyperref}
\usepackage{caption}
\usepackage{authblk}
\usepackage{cite}

\hypersetup{colorlinks=true,linkcolor=blue,citecolor=blue,urlcolor=blue}

\begin{document}
\title{Strategic OTC market making with reputation feedback}
\author{Alexander \textsc{Barzykin}\footnote{HSBC, 8 Canada Square, Canary Wharf, London E14 5HQ, United Kingdom, \texttt{alexander.barzykin@hsbc.com}.}}
\date{\today}

\maketitle

\begin{abstract}
Electronic over-the-counter (OTC) liquidity provision is increasingly shaped not only by the price of the next quote, but also by a dealer's accumulated standing with clients and platforms. We develop a stochastic-control model in which request-for-quote (RFQ) win ratios and streaming fill ratios feed back into future flow through performance-based flow gates, creating an explicit trade-off between immediate spread capture and long-term franchise value. The resulting policy naturally alternates between reputation-building campaigns and franchise monetization phases, and can generate multiple stable client-flow regimes even in a parsimonious single dealer control problem.
\end{abstract}

\vspace{7mm}

\textbf{Key words:} Market Making, Stochastic Optimal Control, Algorithmic Trading, Last Look, Adverse Selection, Latency, Win Ratio, Fill Ratio, Bifurcation, Bistability.

\vspace{5mm}

\section*{Introduction}

In electronic OTC markets, a dealer's problem is no longer just a sequence of independent quote optimizations. RFQ platforms, aggregators and streaming channels increasingly allocate attention and flow according to recent execution quality: competitiveness, rejection behaviour, latency experience, markouts and the perceived reliability of the relationship. From a business perspective, this turns liquidity provision into a dynamic franchise problem. A dealer who loses too many RFQs may cease to be invited to provide a price, or may receive only low-quality opportunities; a streaming dealer who rejects too often may be down-weighted by clients or aggregators; and a dealer with strong recent performance can obtain leadership in a client segment, creating more frequent opportunities but also greater inventory and toxicity exposure.

Classical stochastic market making models focus on the trade-off between spread capture and inventory risk, with exogenous request intensities and fill probabilities \cite{AvellanedaStoikov2008,GueantLehalleFernandezTapia2013,CarteaJaimungalRicci2014,BergaultGueant2021}. Central importance of latency and informational risk in OTC environments has inspired a range of studies on last look, adverse selection and price reading
\cite{OomenAggregator2017,OomenLastLook2017,CarteaJaimungalWalton2019,BarzykinBergaultGueantLemmel2025,CarteaSanchez2025}. Recent work has also emphasized explicit dealer competition, broker selection, platform design and learning in dealer-to-client markets \cite{BankEkrenMuhleKarbe2021,Wang2023,BoyceHerdegenSanchez2024,EislerMuhleKarbe2024,ContXiong2024,MarinArdanzaSabio2025,BoyceNeuman2026}. Yet most tractable control models still treat access to future flow as exogenous, whereas in practice access is itself a state variable influenced by today's quoting and acceptance decisions.

This paper proposes a compact framework for this feedback channel. The dealer services two electronic protocols. Tier $A$ represents RFQ flow, where the relevant long-run score is the dealer's win ratio \cite{BarzykinWinRatio2026}. Tier $B$ represents streaming flow, where the relevant score is the fill ratio generated by the trade acceptance protocol. Each score is updated as an exponentially weighted success measure, and future request intensity is multiplied by a score-dependent gate. The dealer therefore controls not only the economics of the next trade, but also the evolution of the future opportunity set. In the streaming tier, adverse-selection risk is modeled through a Gaussian latency mark and a fair acceptance/rebate protocol, so that the dealer jointly optimizes quotes and an option-like last-look threshold \cite{BarzykinLastLook2026}.

Mathematically, the model leads to a Hamilton-Jacobi-Bellman (HJB) equation with two slow reputation variables and a fast inventory state. We use a slow--fast, or adiabatic, approximation: for frozen reputation, the dealer solves the fast inventory-control problem; the induced policy is then averaged over the stationary inventory distribution to obtain a deterministic reputation drift. This reduction preserves the economic feedback loop while making the two-dimensional score dynamics transparent. The numerical results show that performance-based flow gates can create endogenous reputation-building campaigns and franchise monetization phases, cross-tier spillovers between RFQ and streaming policies, and bistable regimes in which a dealer can either remain marginal or become a leading liquidity provider depending on initial reputation and the path of recovery.

\section*{Model formulation}

Let $S_t$ denote the reference mid price and assume that it follows the Brownian motion with constant volatility $\sigma > 0$ on a finite operational horizon $[0, T]$. The dealer services two execution protocols: RFQ and streaming, denoted by tiers $A$ and $B$, respectively. Tier $\tau \in \{A, B\}$ trades with size $z_\tau \in \mathbb Z_+$.\footnote{This assumption is aimed to simplify the presentation of the main effects related to reputation feedback. Generalization to a ladder of sizes is well understood \cite{BergaultGueant2021}.} The dealer controls the quote offsets $\delta^{\tau,s}_t$ per side $s \in \{b,a\}$ (bid/ask), so the executable quotes are
\[
S_t^{\tau, s} = S_t - \epsilon_s \delta^{\tau, s}_t,\qquad \epsilon_b = +1, \quad \epsilon_a = -1.
\]
For tier $B$ (streaming), the dealer also controls a nonnegative rejection threshold $\eta^{B,s}_t$.

The two tiers differ in the economically relevant event clock. For each side $s\in\{b,a\}$, let $(M_t^{A,s})_{t\ge 0}$ denote the counting process of observed quote requests in tier $A$. Its predictable intensity is $\lambda^{A,s}G_A(R_t^A)$, where $\lambda^{A,s}>0$ is the baseline RFQ intensity and $G_A$ is the gate function of the RFQ score $R^A$ (defined below). Upon observing a request, the dealer quotes $\delta_t^{A,s}$. The RFQ is then won with probability $f^{A,s}(\delta_t^{A,s})\in[0,1]$. If the RFQ is won, a trade of size $z_A$ takes place, no trade occurs otherwise. Thus the fill counting process $N^{A,s}$ is a controlled thinning of $M^{A,s}$.

In the streaming tier $B$, the dealer does not observe a preliminary request for quote. The relevant observed events are directly the trade requests against the stream. For each side $s\in\{b,a\}$, let $(M_t^{B,s})_{t\ge 0}$ denote the counting process of observed trade requests. Its predictable intensity is $\lambda^{B,s}G_B(R_t^B)f^{B,s}(\delta_t^{B,s})$, where $\lambda^{B,s}>0$ is the baseline streaming trade-request intensity, $G_B$ is the gate function of the streaming score $R^B$, and $f^{B,s}$ captures the sensitivity of observed trade-request flow to the quoted stream.

Due to latency, the dealer's quote at the time of trade request arrival can be different from the quote to which the client responded. Let $Y_\tau \sim \mathcal N(m_\tau, \nu_\tau^2)$ denote the sign-adjusted latency mark of trade requests. The mean $m_\tau < 0$ is generally negative (against the dealer) due to adverse selection (winner's curse \cite{OomenAggregator2017}). The dealer can manage toxicity by implementing a trade acceptance protocol. Business-wise, rejections are more common for the streaming flow, so we ignore RFQ rejections here for simplicity.\footnote{
Post-trade adverse selection effects are also out of scope, although important in practice \cite{BarzykinBergaultGueantLemmel2025, CarteaSanchez2025}.
} For the streaming tier, the fair protocol with threshold $\eta^{B,s}_t \ge 0$ reads \cite{OomenLastLook2017}
\[
\begin{cases}
Y_B < -\eta^{B,s}_t, & \text{reject}\\
-\eta^{B,s}_t \le Y_B \le \eta^{B,s}_t, & \text{fill at realized slippage}\\
Y_B > \eta^{B,s}_t, & \text{fill and cap the effective slippage at $\eta^{B,s}_t$}.
\end{cases}
\]
Hence the fill counting process $N^{B,s}$ is a controlled thinning of $M^{B,s}$ generated by the above fair protocol.

\paragraph{Option interpretation of the streaming protocol.}
From a derivatives perspective, the tier-$B$ streaming protocol embeds a barrier-like payoff in the execution mechanism. Conditional on a trade request, the short-horizon slippage mark $Y_B$ plays the role of the underlying. If $Y_B<-\eta$, the trade is knocked out by rejection; if $-\eta\le Y_B\le \eta$, the dealer receives the linear payoff associated with the realized slippage; and if $Y_B>\eta$, the dealer's upside is capped by the fair-protocol rebate. Thus the dealer jointly controls a quote offset $\delta$ and a barrier/cap level $\eta$. This option-like structure gives the streaming Hamiltonian a familiar derivatives flavour, while keeping the protocol directly tied to last-look risk control \cite{BarzykinLastLook2026}.

\paragraph{Inventory and cash.}
For accepted trades, we denote $\hat Y_\tau$ the effective sign-adjusted slippage entering the transaction price: $\hat Y_A = Y_A$, $\hat Y_B = \min(Y_B, \eta^{B,s}_t)$. Thus the accepted trade in tier $\tau$ on side $s$ is executed at price $S_t - \epsilon_s (\delta^{\tau,s}_t + \hat Y_\tau)$. Let $Q_t \in \mathbb Z$ denote inventory and $X_t \in \mathbb R$ cash. Their dynamics are
\[
\mathrm{d} Q_t = \sum_{\tau \in \{A,B\}} z_\tau \left( 
\mathrm{d} N_t^{\tau,b} -\mathrm{d} N_t^{\tau,a} 
\right) ,
\]
\[
\mathrm{d} X_t
= \sum_{\tau \in \{A,B\}} \sum_{s \in \{b,a\}} \epsilon z_\tau
\left(S_t-\epsilon_s \bigl(\delta_t^{\tau,s} + \hat Y_\tau\bigr)\right)\,\mathrm{d} N_t^{\tau,s}.
\]
With this sign convention, the mark-to-market increment $X_t + Q_tS_t$ associated with an accepted trade in tier $\tau$ is $z_\tau (\delta^{\tau,s}_t + \hat Y_\tau)$.

\paragraph{Score dynamics.}
The score is a long-term average measure of success -- win ratio for tier $A$ and fill ratio for tier $B$. For each tier $\tau \in \{A, B\}$, the score update is
\begin{equation}
R^\tau \mapsto (1-\alpha_\tau)R^\tau+\alpha_\tau I^\tau,
\qquad \alpha_\tau\in(0,1],
\end{equation}
where $I^\tau\in\{0,1\}$ is the success indicator attached to one request and $\alpha_\tau \in (0, 1]$ is the memory coefficient (typically $\ll 1$). 
In tier $A$, one observed request is an RFQ. Thus $I^A=1$ if the dealer wins the RFQ and trades, and $I^A=0$ otherwise. In tier B, one observed request is a trade request against the stream. Thus $I^B=1$ if the request is filled and $I^B=0$ if it is rejected. It is convenient to introduce the post-trade score states
\[
R_+^A=\bigl((1-\alpha_A)R^A+\alpha_A,\;R^B\bigr),
\qquad
R_-^A=\bigl((1-\alpha_A)R^A,\;R^B\bigr),
\]
and
\[
R_+^B=\bigl(R^A,\;(1-\alpha_B)R^B+\alpha_B\bigr),
\qquad
R_-^B=\bigl(R^A,\;(1-\alpha_B)R^B\bigr).
\]
The overall dealer score is two-dimensional, $R = (R^A, R^B) \in \mathbb R^2$.

\paragraph{Objective and reduced value function.}
Let $\ell: \mathbb Z \to \mathbb R_+$ be the terminal inventory penalty. The dealer aims to maximize expected terminal mark-to-market wealth penalized by running and terminal inventory risk \cite{CarteaJaimungalRicci2014}
\begin{equation}
V(t,x,q,S,R) := \sup_{\delta,\eta} \mathbb E_{t,x,q,S,R}
\left[ X_T + q_T S_T - \ell(q_T) - \frac12\gamma\sigma^2 \int_t^T  q_u^2\,\mathrm{d} u \right],
\end{equation}
where $\gamma > 0$ is the risk aversion coefficient. Controls are quote offsets and rejection threshold. Because the objective is linear in cash, we introduce the reduced value function $\theta$ through standard ansatz 
\[
V(t,x,q,S,R)=x+qS+\theta(t,q,R).
\]

\section*{HJB equation and optimal controls}

It is convenient to define the post-request value function increments
\begin{equation}
\Delta_-^\tau\theta(t,q,R) := \theta(t,q,R_-^\tau)-\theta(t,q,R),
\end{equation}
\begin{equation}
\Delta^{\tau,s}_+ \theta(t,q,R) :=
\theta\bigl(t,q+\epsilon_s z_\tau,R_+^\tau\bigr)-\theta(t,q,R),
\end{equation}
and reduced state variables
\begin{equation}
\xi_\tau(t,q,R):=\frac{\Delta_-^\tau\theta(t,q,R)}{z_\tau},
\end{equation}
\begin{equation}
p_\tau^s(t,q,R) :=
\frac{\theta(t,q,R_-^\tau)-\theta\bigl(t,q+\epsilon_s z_\tau,R_+^\tau\bigr)}{z_\tau},
\end{equation}
and write the HJB equation in the following form
\begin{equation}
0=\partial_t\theta-\frac12\gamma\sigma^2 q^2
+\sum_{\tau, s}\lambda^{\tau,s}G_\tau(R^\tau)\,z_\tau H_{\tau}^s\bigl(\xi_\tau,p_\tau^s\bigr),
\end{equation}
with terminal condition $\theta(T,q,R) = -\ell(q)$, where $H^s_\tau(\xi, p)$ are the tier-specific Hamiltonians to be defined below.

\paragraph{RFQ contribution.}
Here and below, the contribution of one observed request means the conditional term that enters the reduced HJB generator: the expected immediate mark-to-market gain plus the corresponding continuation-value increment, before multiplication by the baseline arrival intensity. Every observed RFQ in tier $A$ leads either to a loss of the RFQ, with continuation increment $\Delta_-^A\theta$, or to a win, with continuation increment $\Delta_+^{A,s}\theta$ and expected trade gain $z_A(\delta+m_A)$. Here $m_A=\mathbb E[Y_A]$ is the mean RFQ slippage mark, so $z_A(\delta+m_A)$ is the conditional expected mark-to-market gain of a won RFQ. Therefore, the expected contribution of one RFQ request is
\[
(1-f^{A,s}(\delta))\Delta_-^A\theta +
f^{A,s}(\delta)\Bigl(z_A(\delta+m_A)+\Delta_+^{A,s}\theta\Bigr),
\]
which can be rewritten as
\[
\Delta_-^A\theta+z_A f^{A,s}(\delta)(\delta+m_A-p_A^s).
\]
Hence the tier $A$ Hamiltonian can be cast into
\begin{equation}
H_A^s(\xi_A, p^s_A):= \xi_A + \mathcal H^s_A(p^s_A - m_A),
\end{equation}
where
\begin{equation}
\mathcal H^s_A(p) := \sup_{\delta\in\mathbb R} f^{A,s}(\delta)(\delta-p).
\end{equation}

\paragraph{Streaming contribution.}
The same convention is used for the streaming tier: the contribution of one observed trade request is the expected immediate mark-to-market gain, including the fair-protocol slippage/rebate term, plus the continuation-value increment associated with the resulting score update. Every observed tier $B$ trade request updates the score, either to $R_-^B$ if the request is rejected or to $R_+^B$ if it is filled. For fixed $(\delta,\eta)$, this one-request contribution is therefore
\[
\begin{aligned}
&\mathbb E\Bigl[
\Delta_-^B\theta\,\mathbbm 1_{\{Y_B< -\eta\}}
+\bigl(z_B(\delta+\min(Y_B,\eta))+\Delta_+^{B,s}\theta\bigr)\mathbbm 1_{\{Y_B\ge -\eta\}}\Bigr]\\
&\hspace{5.4em}
= z_B\big(\xi_B + K_B(\delta, \eta, p^s_B)\big),
\end{aligned}
\]
where the fair-protocol kernel is defined as
\begin{equation}
K_B(\delta,\eta,p):=
\mathbb E\Bigl[
\big(\delta+\min(Y_B,\eta)-p\big)\mathbbm 1_{\{Y_B\ge -\eta\}}
\Bigr] = \beta_B(\eta)(\delta-p) + L_B(\eta) ,
\end{equation}
with 
\begin{equation}
\beta_B(\eta) := 1-\Phi^-, \qquad L_B(\eta) := m_B \big( \Phi^+ - \Phi^- \big) + \nu_B \big( \phi^- - \phi^+ \big) + \eta \big( 1 - \Phi^+ \big),
\end{equation}
\begin{equation}
\Phi^\pm := \Phi\left(\frac{\pm \eta-m_B}{\nu_B}\right), \quad \phi^\pm := \phi\left(\frac{\pm \eta-m_B}{\nu_B}\right),
\end{equation}
where $\phi$ and $\Phi$ denote the standard normal density and distribution function.
Accordingly, the streaming Hamiltonian is given by
\begin{equation}
H_B^s(\xi,p):=
\sup_{\delta\in\mathbb R,\,\eta\ge 0} f^{B,s}(\delta)\bigl(\xi+K_B(\delta,\eta,p)\bigr).
\end{equation}

\paragraph{Optimal controls.}
Once the reduced value function $\theta$ is known, the optimal controls are recovered pointwise from the Hamiltonians. For tier $A$, the optimal quote satisfies
\begin{equation}
\delta_A^{*,s}(t,q,R) = \arg\max_{\delta\in\mathbb R} f^{A,s}(\delta)\bigl(\delta + m_A - p_A^s(t,q,R)\bigr).
\label{eq:A_control}
\end{equation}
If $H_A^s$ is differentiable and the maximizer is unique, the envelope theorem can be used to express the optimal quote via the inverse function $(f^A,s)^{-1}$ \cite{BergaultGueant2021}.

For tier $B$, the optimal quote and fair threshold are obtained jointly from
\begin{equation}
(\delta_B^{*,s},\eta_B^{*,s})(t,q,R) = \arg\max_{\delta\in\mathbb R,\,\eta\ge 0}
 f^{B,s}(\delta)\bigl(\xi_B(t,q,R)+K_B(\delta,\eta,p_B^s(t,q,R))\bigr).
 \label{eq:B_control}
\end{equation}
If the maximizer is unique and interior, it can be expressed via a system of two first-order conditions.

\section*{Adiabatic approximation}

The main structural feature of the model is the separation between the fast time scale of inventory risk management and the slow time scale of score change. Quotes and acceptance thresholds are updated request by request, whereas the score variables $R^\tau$ move only through increments of size $\alpha_\tau \ll 1$. This invites an adiabatic treatment in which the score is frozen when solving the fast inventory problem and then evolved on a slower time scale using the induced success rates. We retain first-order terms in $\alpha_\tau$ for both tiers. For smooth $\theta$, the score-shifted values satisfy, for $\tau \in \{A,B\}$,
\begin{equation}
\theta(t,q,R_+^\tau) = \theta(t,q,R) +
\alpha_\tau(1 -R^\tau)\,\partial_{R^\tau}\theta(t,q,R) + O(\alpha_\tau^2),
\label{eq:theta_R_plus}
\end{equation}
\begin{equation}
\theta(t,q,R_-^\tau) = \theta(t,q,R) -
\alpha_\tau R^\tau\,\partial_{R^\tau}\theta(t,q,R) + O(\alpha_\tau^2).
\label{eq:theta_R_minus}
\end{equation}
Consequently,
\begin{equation}
\xi_\tau(t,q,R) = -\alpha_\tau \frac{R^\tau}{z_\tau} \partial_{R^\tau}\theta(t,q,R) + O(\alpha_\tau^2).
\label{eq:xi_tau_expansion}
\end{equation}
\begin{equation}
p_\tau^s(t,q,R) = p_{\tau,0}^s(t,q,R) - \alpha_\tau \chi_\tau^s(t,q,R) + O(\alpha_\tau^2),
\label{eq:p_tau_expansion}
\end{equation}
where $p_{\tau,0}^s(t,q,R) := \left. p_\tau^s(t,q,R)\right|_{\alpha_\tau=0}$ and
\begin{equation}
\chi_\tau^s(t,q,R) := 
\frac{1}{z_\tau} \left[R^\tau\,\partial_{R^\tau}\theta(t,q,R) + (1-R^\tau)\,\partial_{R^\tau}\theta(t,q+\epsilon_s z_\tau,R)\right].
\label{eq:chi_tau_definition}
\end{equation}
Using these notations, we can expand the Hamiltonians
\begin{equation}
H_\tau^s\bigl(\xi_\tau,p_\tau^s\bigr) = H_\tau^s\bigl(0,p_{\tau,0}^s\bigr) -
\alpha_\tau
\left[
\frac{R^\tau}{z_\tau}\partial_{R^\tau}\theta\,
\partial_\xi H_\tau^s\bigl(0,p_{\tau,0}^s\bigr) + \chi_\tau^s\,\partial_p H_\tau^s\bigl(0,p_{\tau,0}^s\bigr) \right] + O(\alpha_\tau^2),
\label{eq:H_adiabatic_expansion}
\end{equation}
where all functions are evaluated at $(t, q, R)$. Note that $\partial_\xi H_A^s(0, p_{A,0}^s) = 1$.

\paragraph{Frozen-score ergodic fast problem.}
For each fixed score vector $R$, we consider the long-horizon fast inventory problem. It is useful to separate the slowly varying score value from the stationary inventory corrector and write
\begin{equation}
\theta(t,q,R)
\approx
U(t,R)+\vartheta^R(q),
\label{eq:slow_fast_decomposition}
\end{equation}
where $U$ captures the slow score value and $\vartheta^R$ is the frozen-score stationary corrector. Substituting this decomposition into the definitions of $p_{\tau,0}^s$ and $\chi_\tau^s$ separates the two roles of the approximation: the inventory dependence of the fast controls is encoded in $\vartheta^R$, while the strategic value of changing the score enters through the score gradients $\partial_{R^\tau}U$ and $\partial_{R^\tau}\vartheta^R$. In particular, at fixed $R$ we obtain
\begin{equation}
p_{\tau,0}^{s,R}(q) = \frac{\vartheta^R(q)-\vartheta^R(q+\epsilon_s z_\tau)}{z_\tau},
\label{eq:p0_stationary}
\end{equation}
\begin{equation}
\chi_\tau^{s,R}(q) = \frac{R^\tau D_\tau^R(q) + (1-R^\tau)D_\tau^R(q+\epsilon_s z_\tau)}{z_\tau},
\label{eq:chi_stationary}
\end{equation}
where
\begin{equation}
D_\tau^R(q) := \partial_{R^\tau}U(t,R)+\partial_{R^\tau}\vartheta^R(q).
\label{eq:score_gradient_decomposition}
\end{equation}
This notation keeps the strategic score gradient explicit without committing to a particular terminal horizon or discounting convention for the slow score problem.

\paragraph{Quadratic BEGV approximation of the fast problem.}

To obtain explicit controls, we combine the BEGV quadratic Hamiltonian expansion with a quadratic ansatz for the value function \cite{BEGV2021}. We assume, for this approximation, side symmetry within each tier: $\lambda^{\tau,b}=\lambda^{\tau,a}=:\lambda^\tau$,
$H_\tau^b(0,\cdot)=H_\tau^a(0,\cdot)=:H_\tau$. In this case, no linear term in inventory is needed, and we use
\begin{equation}
\vartheta^R(q) \approx -\frac12 A(R)q^2+C(R),
\label{eq:quadratic_corrector}
\end{equation}
and then
\begin{equation}
p_{\tau,0}^{s,R}(q) = \epsilon_s A(R)q+\frac12A(R)z_\tau.
\label{eq:p0_quadratic}
\end{equation}
We expand the frozen-score Hamiltonians around the origin:
\begin{equation}
H_\tau(p) = h_{\tau,0} + h_{\tau,1}p + \frac12 h_{\tau,2}p^2 + O(p^3).
\label{eq:BEGV_hamiltonian_expansion}
\end{equation}
Substituting all the expressions into the HJB equation and matching the coefficients of $q^2$ yields the stationary Riccati relation
\begin{equation}
\frac12\gamma\sigma^2 = A(R)^2
\sum_\tau
\lambda^\tau G_\tau(R^\tau)z_\tau h_{\tau,2} + O(\alpha_A+\alpha_B).
\label{eq:stationary_Riccati}
\end{equation}
This equation determines the frozen-score inventory slope $A(R)$ at leading order.

The first-order score-gradient correction to shadow price used in controls is obtained as
\begin{equation}
p_\tau^s(q,R) \approx p_{\tau,0}^{s,R}(q)-\alpha_\tau\chi_\tau^{s,R}(q),
\label{eq:corrected_shadow_price_controls}
\end{equation}
where
\begin{equation}
D_\tau^R(q) \approx \partial_{R^\tau}U(t,R) -\frac12 A_{R^\tau}(R)q^2+C_{R^\tau}(R).
\label{eq:D_quadratic}
\end{equation}
The controls are obtained from the exact point-wise maximizers \eqref{eq:A_control} and \eqref{eq:B_control}.

\paragraph{Slow score dynamics.}
For a fixed frozen-score feedback policy, let \(\mu^R\) denote the invariant law of the corresponding fast inventory dynamics. 
Averaging the score updates with respect to \(\mu^R\) yields the effective slow drift of the reputation state. 
For the RFQ tier,
\begin{equation}
\dot R^A = \alpha_A \sum_{s\in\{b,a\}} \lambda^{A,s}G_A(R^A)
\int \left( f^{A,s}\bigl(\delta_A^{*,s}(q,R)\bigr)-R^A \right) \mu^R(\mathrm d q),
\label{eq:slow_RA}
\end{equation}
whereas for the streaming tier,
\begin{equation}
\beta_B^s(q,R)
=
\mathbb P\!\left(Y_B\ge -\eta_B^{*,s}(q,R)\right)
=
1-\Phi\!\left(\frac{-\eta_B^{*,s}(q,R)-m_B}{\nu_B}\right),
\label{eq:betaB}
\end{equation}
and
\begin{equation}
\dot R^B = \alpha_B \sum_{s\in\{b,a\}} \lambda^{B,s}G_B(R^B)
\int f^{B,s}\bigl(\delta_B^{*,s}(q,R)\bigr)\left(\beta_B^s(q,R)-R^B\right)\mu^R(\mathrm d q).
\label{eq:slow_RB}
\end{equation}

\paragraph{Reduced stationary score problem.}
For the numerical computation of the continuation value, we replace the finite-horizon slow equation by a discounted stationary problem on the score space. 
Given a frozen-score policy \(\pi\), the invariant law \(\mu^R\) determines both the inventory-averaged objective rate \(\bar r^\pi(R)\) and the effective score-jump intensities, thereby inducing a controlled pure-jump Markov chain with generator \(\mathcal L_R^\pi\). 
The associated reduced value is defined by
\[
\rho U^\pi(R)=\bar r^\pi(R)+(\mathcal L_R^\pi U^\pi)(R).
\]
The numerical scheme is then based on policy iteration for this discounted stationary problem, with exact policy evaluation at each step. 
Thus, the reduced value \(U\) entering the controls is obtained from a stationary discounted approximation, while \eqref{eq:slow_RA}--\eqref{eq:slow_RB} describe the deterministic slow drift induced by the resulting policy.

\section*{Numerical results}
\label{sec:numerical-results}

We use a truncated inventory grid and a rectangular score grid. Off-grid score values are evaluated by bilinear interpolation. For the RFQ tier we use a sigmoid trade probability
\begin{equation}
 f_A(\delta)=\bigl(1+\exp(a_A+b_A\delta)\bigr)^{-1}.
\end{equation}
At each grid point $(q,R)$ and side $s$, the optimal RFQ offset solves \eqref{eq:A_control}. Writing $c_A^s(q,R)=p_A^s(q,R)-m_A$, the closed-form solution is
\begin{equation}
\delta_A^{*,s}(q,R) = c_A^s(q,R) +
\frac{1}{b_A} \Big(1+W\!\bigl[\exp\{-1-a_A-b_Ac_A^s(q,R)\}\bigr]\Big),
\label{eq:numerics_deltaA_lambert}
\end{equation}
where $W$ is the Lambert function.

For the streaming tier we use the exponential trade-request shape
\begin{equation}
 f_B(\delta)=\exp(-\kappa_B\delta).
\end{equation}
For a fixed threshold $\eta$, the streaming objective is
\[
\exp(-\kappa_B\delta)
\left(\xi_B+\beta_B(\eta)(\delta-p_B^s)+L_B(\eta)\right),
\]
so the optimal offset conditional on $\eta$ is
\begin{equation}
\delta_B^{*,s}(q,R;\eta) = \frac{1}{\kappa_B} -
\frac{\xi_B(q,R)-\beta_B(\eta)p_B^s(q,R)+L_B(\eta)}{\beta_B(\eta)}.
\label{eq:numerics_deltaB_conditional}
\end{equation}
The remaining one-dimensional optimization over $\eta\ge0$ is performed on a fine grid.  This gives $(\delta_B^{*,s},\eta_B^{*,s})$ at every $(q,R)$.

The score gate function is assumed to be logistic for both tiers:
\begin{equation}
G_\tau(R^\tau) = G_\tau^{\min} + \big(1 - G_\tau^{\min}\big) \left( 1 + e^{-\upsilon_\tau (R^\tau - R_0^\tau)}\right)^{-1}, 
\end{equation}
with steepness $\upsilon_\tau$ and midpoint $R_0^\tau$.

Default parameters for the two tiers are reported in Table~\ref{tab:params}. They are used throughout unless explicitly stated otherwise. In addition, we assume daily volatility of 100 bp and a risk aversion coefficient of $\gamma = 5\cdot 10^{-4}$ bp$^{-1}$ M$^{-1}$. The reduced value function is computed by damped policy iteration with discount $\rho = 0.05$ and exact sparse policy evaluation for each fixed policy. Unless stated otherwise, the displayed control surfaces are evaluated at zero inventory, $q=0$, while reduced drifts and success probabilities are averaged with respect to the frozen-score stationary inventory law. The inventory grid was restricted to $\pm 50$ M, the score grid has size $100 \times 100$, and both the threshold an spread grids contain 100 points. The tier $B$ total spread is constrained from below to 0.3 bp to avoid economically meaningless streaming quotes.

\begin{table}[h!]
\centering
 \begin{tabular}{c c c c c c c c c c} 
 \hline
 Tier & $z_\tau$ & $\lambda^{\tau,s}$ & $f_\tau(\delta)$ & $m_\tau$ & $\nu_\tau$ & $G_\tau^{\min}$ & $R_0^\tau$ & $\upsilon_\tau$ & $\alpha_\tau$ \\ 
 \hline
A & 10 & 100 & $\left(1+\exp(-3.4+10\delta)\right)^{-1}$ & -0.1 & 0.3 & 0.1 & 0.35 & 60 & 0.001 \\
B & 1 & 500 & $\exp(-3\delta)$ & -0.2 & 0.5 & 0.1 & 0.8 & 10 & 0.001 \\
 \hline
 \end{tabular}
\caption{Baseline parameters for the RFQ tier $A$ and the streaming tier $B$. Sizes $z_\tau$ are in million notional, intensities $\lambda^{\tau,s}$ are per day, latency mark parameters $m_\tau$ and $\nu_\tau$ are in basis points (bp), and the slope parameters $b_A$ and $\kappa_B$ and in bp$^{-1}$.
}
\label{tab:params}
\end{table}

\begin{figure}[h!]
    \centering
    \includegraphics[width=0.49\textwidth]{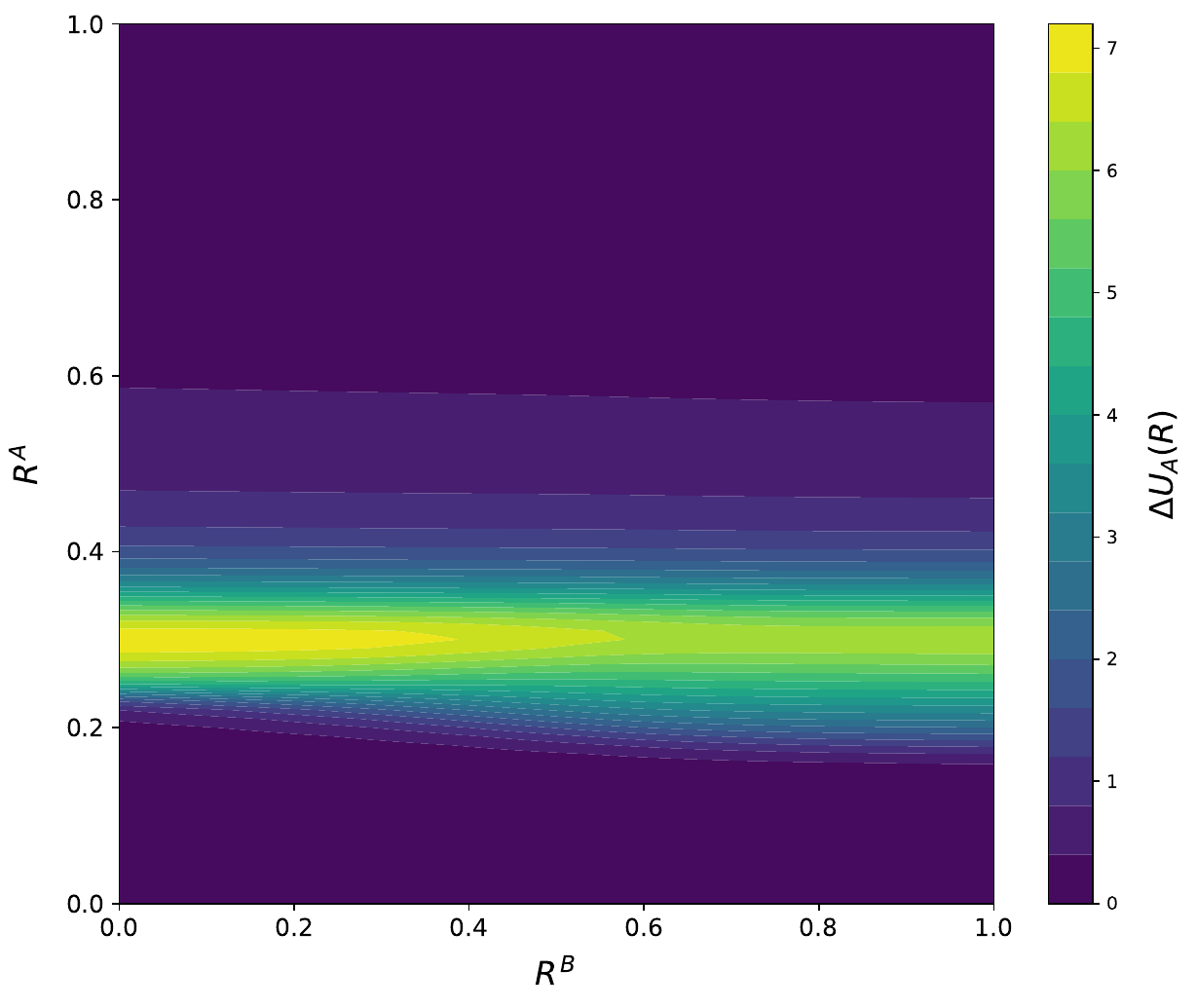}
    \includegraphics[width=0.49\textwidth]{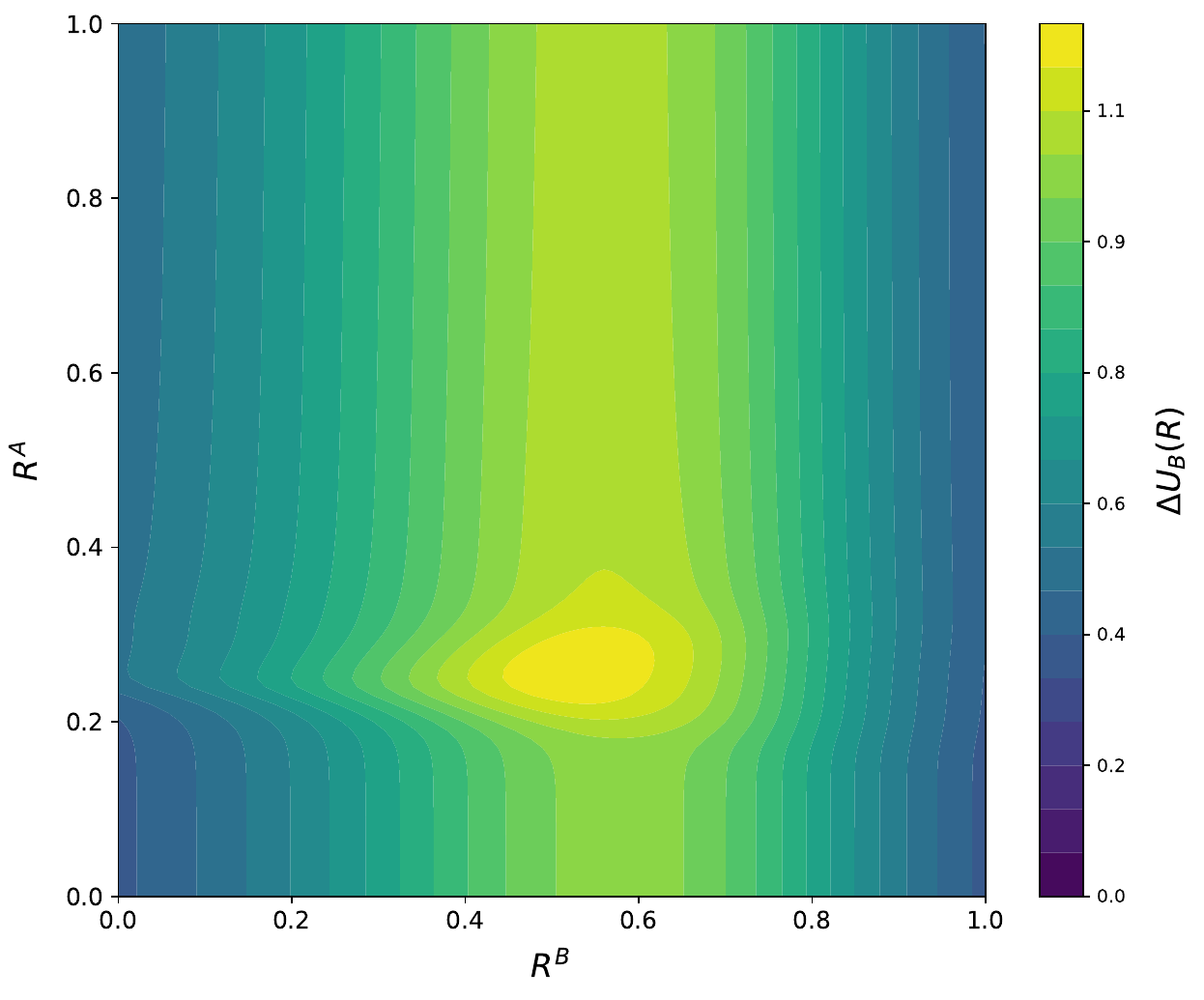}
    \caption{Promotion values $\Delta U_\tau(R) = U(R^\tau_+) - U(R^\tau_-)$ over the two-dimensional score space.}
    \label{fig:promo}
\end{figure}

Figure \ref{fig:promo} displays the marginal continuation value of a successful event in each tier. For tier $A$, this is the continuation value gain of winning rather than losing an RFQ event. The effect is concentrated around the RFQ gate because a single win matters most when the dealer is close to a promotional threshold. For tier~$B$, the promotion value measures the continuation gain from filling rather than rejecting a streaming request. The smoother shape reflects the less abrupt streaming gate and the fact that streaming remains economically active over a wider portion of the state space avoiding rejection-driven decay.

\begin{figure}[h!]
    \centering
    \includegraphics[width=0.49\textwidth]{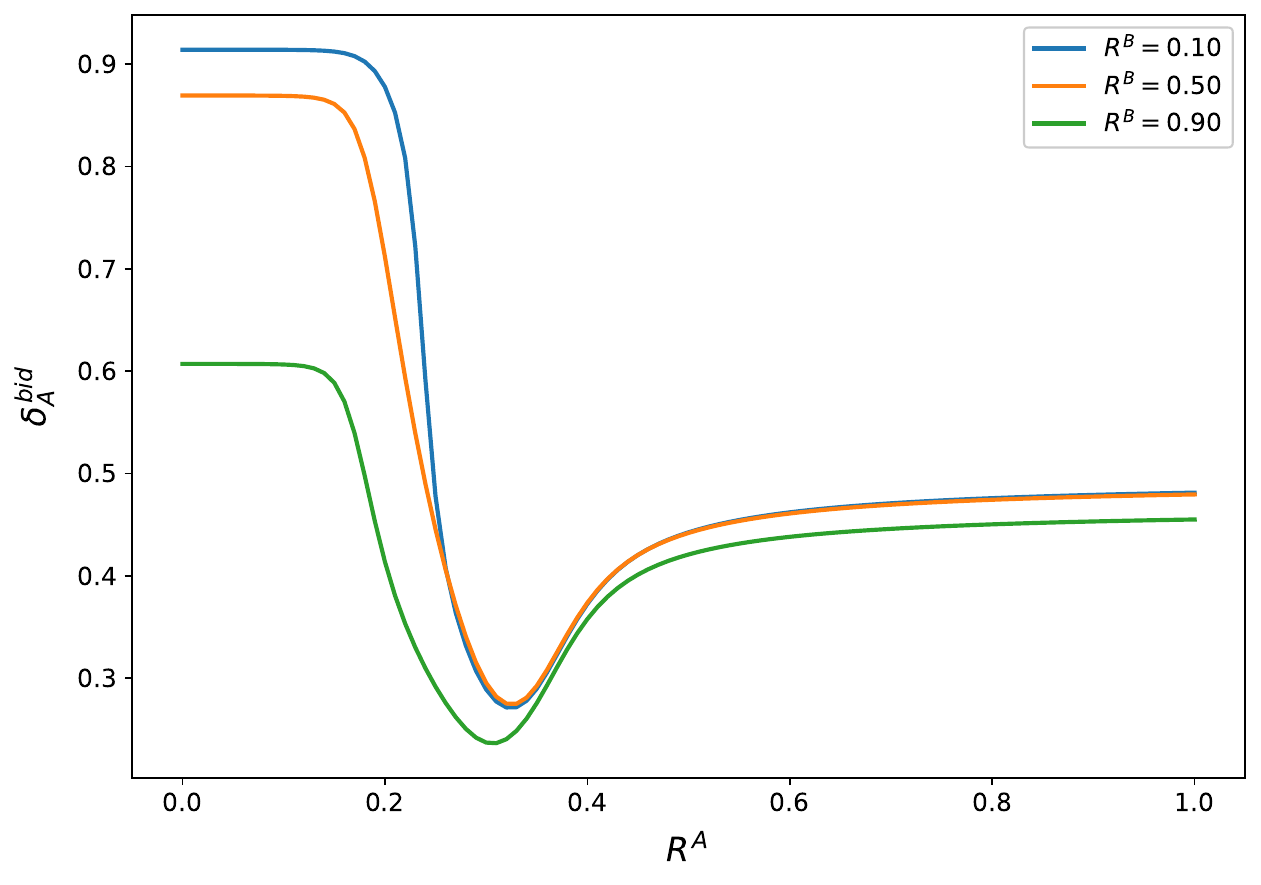}
    \includegraphics[width=0.49\textwidth]{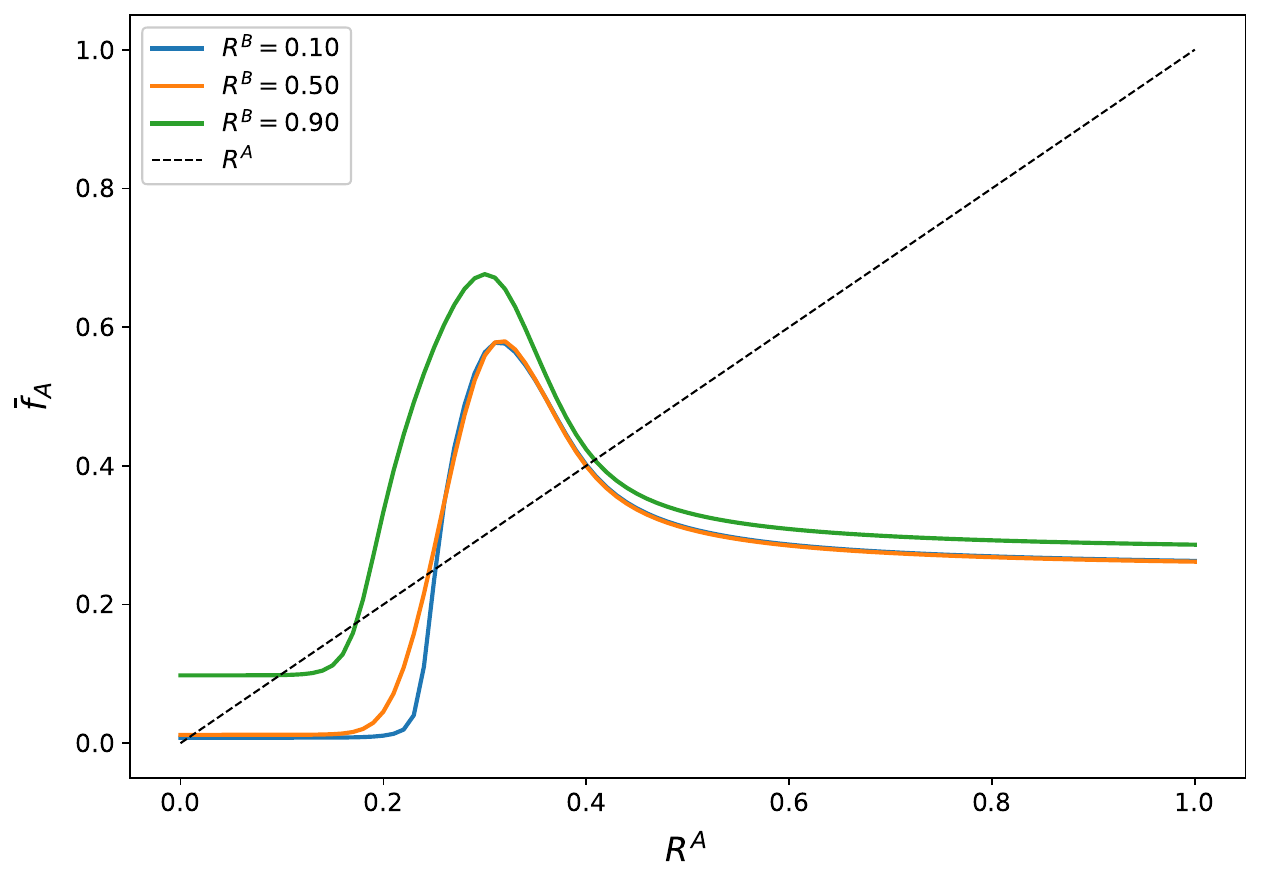}
    \caption{RFQ pricing and score fixed points. Left: bid offset at zero inventory as a function of the RFQ score $R^A$ for three streaming scores $R^B$. Right: inventory-averaged RFQ win probability $\bar f(R)$ against the identity line. Intersections are fixed points of the one-dimensional RFQ score drift for the corresponding $R^B$ slice.}
    \label{fig:tier_A}
\end{figure}

Figure \ref{fig:tier_A} shows how the RFQ tier reacts to reputation. The bid offset falls near the gate midpoint, which is the signature of a campaign: the dealer sacrifices immediate spread to improve the probability of winning and to protect future access. Above the gate, spreads widens as the dealer monetizes the increased flow, although the policy remains defensive because falling back through the gate is costly. The dependence on $R^B$ is moderate but systematic: a stronger streaming franchise improves inventory mixing and can support tighter RFQ prices. The right panel provides a direct visualization of bistability. For the displayed slices, the inventory-averaged win probability $\bar f_A(R)$ curve can intersect the identity line three times, corresponding to low-score and high-score stable equilibria separated by an unstable middle fixed point. The loci of the fixed points depend on the gate contrast and smoothness \cite{BarzykinWinRatio2026}. As $\upsilon$ decreases, a stable + unstable pair annihilates leaving one stable branch.

\begin{figure}[h!]
    \centering
    \includegraphics[width=0.49\textwidth]{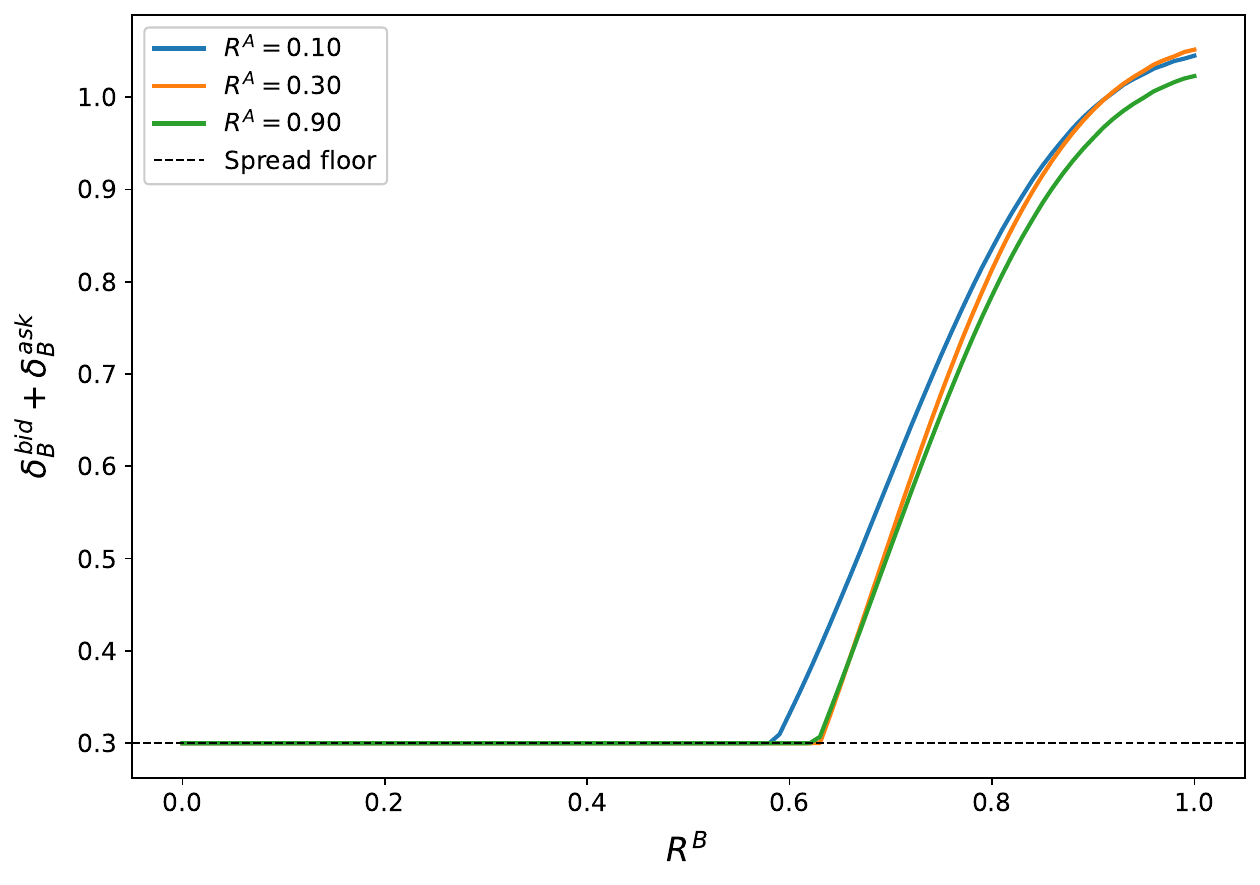}
    \includegraphics[width=0.49\textwidth]{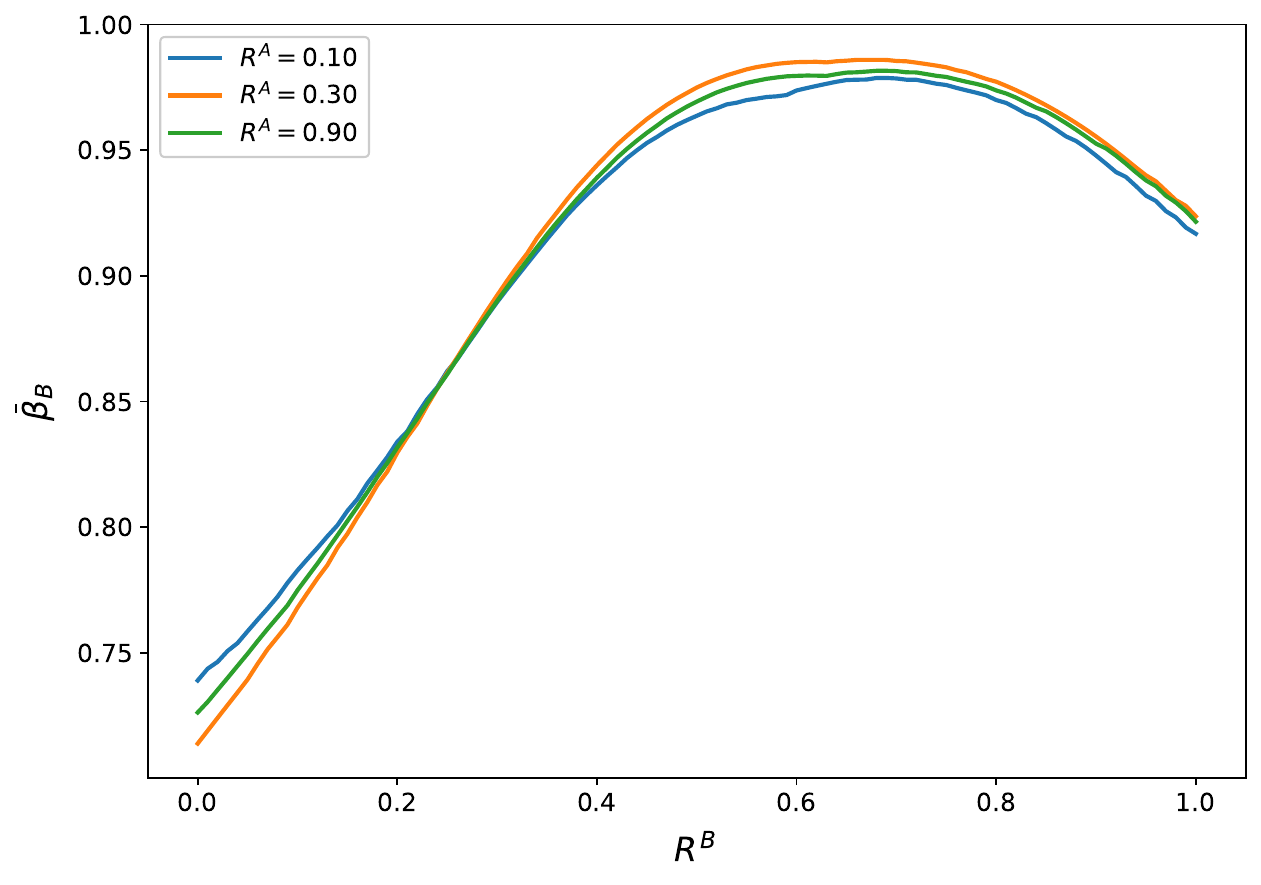}
    \caption{Streaming tier pricing and acceptance policy. Left: total streamed spread $\Delta_B(R,q) = \delta^b_B(R,q) + \delta^a_B(R,q)$ at zero inventory as a function of the streaming score $R^B$, for several RFQ scores $R^A$. Right: inventory-averaged streaming acceptance probability $\bar \beta_B(R)$.}
    \label{fig:tier_B}
\end{figure}

Figure \ref{fig:tier_B} summarizes the streaming response. When $R^B$ is low, the dealer tightens quotes to the spread floor in order to create more streaming requests and hence more chances to repair the fill-ratio score in physical time. The acceptance probability is nevertheless not pushed mechanically to one: the dealer still controls the last-look threshold and retains protection against adverse selection. As $R^B$ improves, the dealer can become more tolerant to flow and can also widen the stream to monetize a stronger franchise. The RFQ score affects the streaming policy through the common inventory and continuation value, although this cross-tier effect is limited under the baseline calibration.

\begin{figure}[h!]
    \centering
    \includegraphics[width=0.69\textwidth]{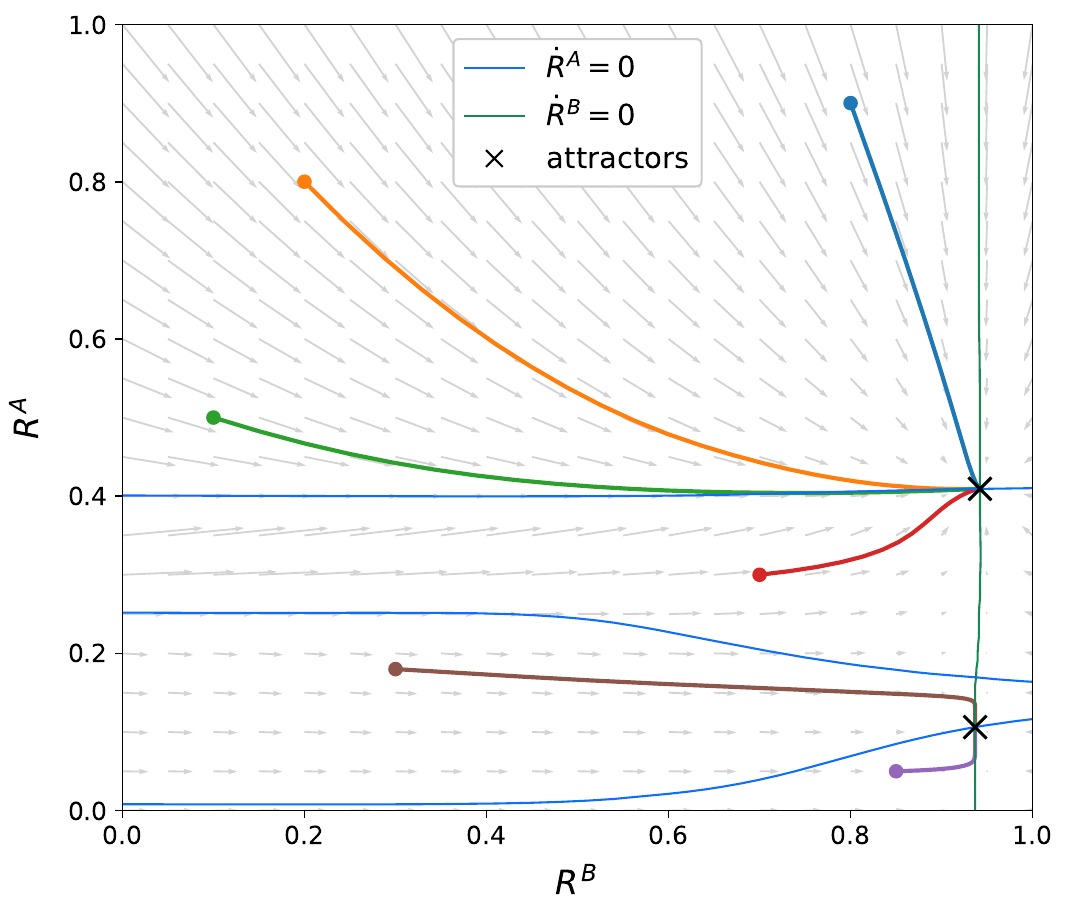}
    \caption{Two-dimensional slow reputation dynamics. Arrows show the averaged score drift, blue and green lines are the nullclines $\dot R^A = 0$ and $\dot R^B=0$, crosses mark attracting equilibria, and colored curves show representative trajectories. The phase portrait exhibits two attraction basins: a low-reputation regime and a high-reputation leadership regime. Folded nullcline geometry also determines the order of recovery across tiers.}
    \label{fig:phase_plot}
\end{figure}

Figure~\ref{fig:phase_plot} combines the two slow score drifts into a phase portrait. In the displayed parameter regime, the system has two attracting equilibria: a low-score state in which the dealer remains marginal, and a high-score state in which the dealer has secured broad client access. The middle branch of the RFQ nullcline separates the two basins, consistent with the fixed-point picture in Figure~\ref{fig:tier_A}. In two dimensions, however, the geometry is richer than a collection of one-dimensional bifurcation diagrams. The RFQ nullcline can fold and form an island-like region without creating an additional attractor. Inside this folded region the vector field is nearly horizontal: the streaming score must recover before the RFQ score can improve. Near the other branch the adjustment order is almost reversed, with the RFQ score relaxing first and the streaming score improving later. Thus the coupled reputation dynamics determine not only the final regime, but also the sequence in which the dealer repairs imbalances across client channels.

\section*{Concluding remarks}

Electronic OTC market making with performance based routing is a dynamic relationship, not only a myopic spread optimization problem. A quote changes current P\&L, inventory risk and adverse-selection exposure, but it also changes the dealer's future opportunity set through the score used by clients and platforms to allocate flow. Once this feedback is included, optimal behaviour naturally separates into campaigning, defense and franchise monetization phases.

The model developed here makes this intuition explicit in a tractable stochastic-control setting. RFQ and streaming channels are represented by two coupled reputation variables: a win-ratio score and a fill-ratio score. The dealer optimizes RFQ quotes, streaming quotes and a fair rejection threshold, while score dependent gates convert recent execution quality into future flow intensity. A slow--fast reduction separates the fast inventory control problem from the slow reputation dynamics and yields an interpretable phase portrait on the score space.

Several conceptual conclusions stand out. First, the marginal value of a win or fill is highly state-dependent; it is largest near promotion threshold, not necessarily when the dealer is already a leader. Second, the same quote can be aggressive or conservative depending on whether it is being used to build reputation, defend access or monetize a strong position. Third, cross-tier interaction effects arise through inventory management and continuation value. Finally, reputation feedback can create multiple stable regimes and path dependence. A dealer with the same technology and risk appetite may remain trapped in a low-flow state or become a leading liquidity provider depending on initial conditions, recovery sequence and the steepness of client-routing gates.

These effects are directly relevant for practical e-trading. They suggest that score management should be evaluated at the portfolio and client-channel level, rather than as a collection of isolated quote decisions. They also provide a modeling language for discussing when to invest in competitiveness, when to tolerate toxic flow, when to protect the franchise through conservative rejection settings, and when to monetize established leadership.

\section*{Acknowledgment}

The author is grateful to Eric Mathew John (HSBC) and Axel Ciceri (HSBC) for fruitful discussions and to Richard Anthony (HSBC) for support throughout the project and valuable comments. The views expressed are those of the author and do not necessarily reflect the views or practices at HSBC.

\end{document}